\documentclass[12pt]{article}
\usepackage{amsfonts,amssymb,amsmath}
\usepackage{graphicx}
\usepackage{dcolumn}
\usepackage{bm}
\usepackage{epsfig}
\usepackage{cite}
\usepackage{color}
\let\a=\alpha \let\b=\beta \let\g=\gamma \let\d=\delta \let\e=\epsilon
  \let\th=\theta  \let\k=\kappa
\let\l=\lambda \let\m=\mu \let\n=\nu \let\x=\xi  
\let\s=\sigma \let\t=\tau    
 
      \let\G=\Gamma  \let\Th=\Theta 
\let\X=\Xi  \let\S=\Sigma  \let\Y=\Psi
 
\let\la=\label  
  
\def\nn{\nonumber} \def\bd{\begin{document}} \def\ed{\end{document}}
\def\ds{\documentstyle} \let\fr=\frac \let\bl=\bigl \let\br=\bigr
\let\Br=\Bigr \let\Bl=\Bigl
\let\bm=\bibitem
\let\na=\nabla
\def\tU{{\widetilde U}}
\let\pa=\partial \let\ov=\overline
\def\ie{{\it i.e.\ }}
\newcommand{\be}{\begin{equation}}
\newcommand{\ee}{\end{equation}}
\def\ba{\begin{array}}
\def\ea{\end{array}}
\def\ft#1#2{{\textstyle{{\scriptstyle #1}\over {\scriptstyle #2}}}}
\def\fft#1#2{{#1 \over #2}}
\def\F#1#2{{ F_{#1}^{(#2)} }}
\def\cF#1#2{{ {\cal F}_{#1}^{(#2)} }}
\def\R{{\bf R}}
\def\sst#1{{\scriptscriptstyle #1}}
\def\oneone{\rlap 1\mkern4mu{\rm l}}
\def\e7{E_{7(+7)}}
\def\td{\tilde}
\def\wtd{\widetilde}
\def\im{{\rm i}}
\def\bog{Bogomol'nyi\ }
\newcommand{\ho}[1]{$\, ^{#1}$}
\newcommand{\hoch}[1]{$\, ^{#1}$}
\newcommand{\bea}{\begin{eqnarray}}
\newcommand{\eea}{\end{eqnarray}}
\newcommand{\ra}{\rightarrow}
\newcommand{\lra}{\longrightarrow}
\newcommand{\Lra}{\Leftrightarrow}
\newcommand{\ap}{\alpha^\prime}
\newcommand{\bp}{\tilde \beta^\prime}
\newcommand{\cB}{{\cal B}}
\newcommand{\cO}{{\cal O}}
\newcommand{\vecx}{\vec{x}}
\newcommand{\vecy}{\vec{y}}
\newcommand{\vecp}{\vec{p}}
\newcommand{\vecq}{\vec{q}}
\newcommand{\tr}{{\rm tr} }
\newcommand{\Tr}{{\rm Tr} }
\newcommand{\NP}{Nucl. Phys. }
\newcommand{\cL}{{\cal L}}
\newcommand{\cA}{{\cal A}}
\newcommand{\cT}{{\cal T}}
\newcommand{\cD}{{\cal D}}
\def\sst#1{{\scriptscriptstyle #1}}
\def\0{{\sst{(0)}}}
\def\1{{\sst{(1)}}}
\def\2{{\sst{(2)}}}
\def\3{{\sst{(3)}}}
\def\4{{\sst{(4)}}}
\def\5{{\sst{(5)}}}
\def\6{{\sst{(6)}}}
\def\7{{\sst{(7)}}}
\def\8{{\sst{(8)}}}
\def\ve{\varepsilon}
\def\vf{\varphi}
\def\F{\Phi}
\def\wg{\wedge}
\def\thb{\bar{\theta}}
\def\Thb{\bar{\Theta}}
\def\barp{\bar{p}}
\def\barq{\bar{q}}
\def\barc{\bar{c}}
\def\bard{\bar{d}}
\def\e{\epsilon}
\def \bi{\bibitem}
\def \la {\label}
\def \l {\lambda}
\def\foot{\footnote}
\def \tl  {{\tilde \l}}
\def \sql {{\sqrt \l}}
\def \adss {$AdS_5 \times S^5$\ }
\newcommand{\rf}[1]{(\ref{#1})}
\def \ov {\over}
\def\th{\theta}
\def\Th{\Theta}
\def\vth{\vartheta}
\def\btheta{{\bar\theta}}
\def\ttheta{{{\tilde\theta}}}
\def\bttheta{{{\bar\ttheta}}}
\def\vth{\vartheta}
\def\ra{\rightarrow}
\def\N{{\cal N}}
\def\F{{\cal F}}
\def\uM{\underline{M}}
\def\uA{\underline{A}}
\def\uN{\underline{N}}
\def\uP{\underline{P}}
\def\ua{\underline{a}}
\def\ub{\underline{b}}
\def\uc{\underline{c}}
\def\ud{\underline{d}}
\def\ue{\underline{e}}
\def\uf{\underline{f}}
\def\ui{\underline{i}}
\def\uj{\underline{j}}
\def\uk{\underline{k}}
\def\ual{\underline{\alpha}}
\def\ube{\underline{\beta}}
\def\um{\underline{m}}
\def\un{\underline{n}}
\def\umu{\underline{\mu}}
\def\unu{\underline{\nu}}
\def\ula{\underline{\l}}
\def\uka{\underline{\k}}
\def\usi{\underline{\s}}
\def\urh{\underline{\r}}
\def\cc{\circ}
\def\eqv{\equiv}
\def\ni{\noindent}
\def\Ep{E^{{}^{(+)}}}
\def\Em{E^{{}^{(-)}}}
\def\Mp{M^{{}^{(+)}}}
\def\Mm{M^{{}^{(-)}}}
\def \ha{{1\ov 2}}
\def\r{\rho}
\def\Y{{\rm Y}}
\def\X{{\rm X}}
\def\tY{\tilde{\rm Y}}
\def\tX{\tilde{\rm X}}
\def\dY{\dot{\rm Y}}
\def\dX{\dot{\rm X}}
\def \J {\mathcal{J}}
\def \del {\partial}
\def\dF{\dot{F}}
\def\dG{\dot{G}}
\def\df{\dot{f}}
\def \E {{\cal E}}
\def \S {{\cal S}}
\def \J {{\cal J}}
\def\ms{\mathcal{S}}
\def\mj{\mathcal{J}}
\def\soj{\fr{\ms}{\mj}}
\def \R {{\bf R}}
\def \om {\omega}
\def \bE {\bar E}
\def \x {{\cal X}}
\def \bi{\bibitem}
\def \la {\label}
\def \l {\lambda}
\def\foot{\footnote}
\def \tl  {{\tilde \l}}
\def \sql {{\sqrt \l}}
\def \adss {$AdS_5 \times S^5$\ }
\def \ov {\over}
\def \varpi {{\rm w}}
\def\thb{\bar{\theta}}
\def\Thb{\bar{\Theta}}
\def\mb{\bar{\m}}
\def\ab{\bar{\a}}
\def\zb{\bar{z}}
\def\psib{\bar{\psi}}
\def\barp{\bar{p}}
\def\barq{\bar{q}}
\def\barc{\bar{c}}
\def\bard{\bar{d}}
\def\e{\epsilon}
\def\wb{\bar{w}}
\def\lb{\bar{\l}}
\def\Jb{\bar{J}}
\def\Nb{\bar{N}}
\def\Zb{\bar{Z}}
\def\pab{\bar{\pa}}
\def\At{\tilde{A}}
\def\Bt{\tilde{B}}
\def\Ct{\tilde{C}}
\def\Dt{\tilde{D}}
\def\Et{\tilde{E}}
\def\Ft{\tilde{F}}
\def\Gt{\tilde{G}}
\def\Ht{\tilde{H}}
\def\Jt{\tilde{J}}
\def\Mt{\tilde{M}}
\def\St{\tilde{S}}
\def\Vt{\tilde{V}}
\def\at{\tilde{a}}
\def\bt{\tilde{b}}
\def\ct{\tilde{c}}
\def\dt{\tilde{d}}
\def\et{\tilde{e}}
\def\ft{\tilde{f}}
\def\gt{\tilde{g}}
\def\mt{\tilde{\mu}}
\def\nt{\tilde{\nu}}
\def\tt{\tilde{t}}
\def\tht{\tilde{\th}}
\def\Tht{\tilde{\Th}}
\def\Phit{\tilde{\Phi}}
\def\phit{\tilde{\phi}}
\def\psit{\tilde{\psi}}
\def\xit{\tilde{\xi}}
\def\vft{\tilde{\vf}}
\def\asth{\hat{*}}
\def\phh{\hat{\phi}}
\def\bA{{\bf A}}
\def\ola{\overleftarrow}
\def\ora{\overrightarrow}
\def\alt{\tilde{\a}}
\def\eh{\hat{e}}
\def\eph{\hat{\e}}
\def\ph{\hat{p}}
\def\alh{\hat{\a}}
\def\beh{\hat{\b}}
\def\gah{\hat{\g}}
\def\Fh{\hat{F}}
\def\muh{\hat{\m}}
\def\nuh{\hat{\n}}
\def\thh{\hat{\th}}
\def\dh{\hat{d}}
\def\ih{\hat{i}}
\def\jh{\hat{j}}
\def\kh{\hat{k}}
\def\deh{\hat{\d}}
\def\wh{\hat{w}}
\def\lah{\hat{\l}}
\def\Ah{\hat{A}}
\def\Ch{\hat{C}}
\def\Omh{\hat{\Omega}}
\def\ps{\rlap{\, /}\;\,p }
\def\ks{\rlap{\, /}\;\,k }
\def\gym{g_{YM}}
\def\adot{\dot{a}}
\def\bdot{\dot{b}}
\def\bpa{\bar{\pa}}
\def\pr{\prime}
\def\ssk{\medskip}
\def\sech{\text{sech}}
\def\csch{\text{csch}}
\def\vn{\vec{n}}
\def\vm{\vec{m}}
\begin{document}
\begin{center}
{\Large\bf Worldsheet one-loop energy correction to IIA Giant Magnon
 \\
 \vspace{0.5cm}
  }
Xiaojian Bai$\,^*$, Bum-Hoon Lee$\,^*$ and I. Y. Park$\,^\dagger$
\\
\vspace{0.3cm}
{\it Center for Quantum Spacetime, Sogang University\\
Shinsu-dong 1, Mapo-gu, 121-742 South Korea$\;^*$ \\
}
\vspace{0.3cm}
{\it Department of Natural and Physical Sciences,
Philander Smith College
                               \\
Little Rock, AR 72202, USA$\;^\dagger$ \\
}
\end{center}
\begin{abstract}
We compute one-loop corrections to the energy of a IIA giant magnon solution in the $AdS_4\times \mathbb{C}P^3$ background by using the standard quantum field theory (QFT) techniques.  The string action is expanded around the solution to the quadratic order in the fluctuation fields. The resulting action has 2D coordinate
dependent-coefficients, a feature that complicates the analysis. The solution contains a worldsheet velocity parameter $v$, and is expanded in terms of the parameter. A perturbative analysis is carried out by treating the $v$-dependent parts as vertices. The
 energy is computed by first putting the system in a box of length $L$ and Fourier-transforming the fields
into the discrete momentum modes. We compare our result with the results obtained by the algebraic
curve method.
\end{abstract}
\newpage
\section{Introduction}
A giant magnon (GM) \cite{Hofman:2006xt} is a relatively simple solution of type II string sigma model whose
dispersion relation has been known with certain precision. In this work we consider a
giant magnon of IIA string theory in $AdS_4 \times CP^3$, and analyze the worldsheet
one-loop energy shift by using the standard QFT fluctuation techniques.
The exact dispersion relation of a magnon in $AdS_4 \times CP^3$ is \cite{Beisert:2005tm}\cite{Gaiotto:2008cg}\cite{Lee:2008ui}
\begin{eqnarray}
\Delta \equiv E-J = \sqrt{\frac{Q^2}{4} + 4 h^2(\lambda) \sin^2 \frac{p}{2} }
\label{dr}
\end{eqnarray}
where $Q$ is the number of the magnons and $\l$ is 't Hooft coupling. ($p$ is a parameter that is related to
the worldsheet velocity $v$ by \rf{vp} below.) The function $h(\lambda)$ can be interpolated
between strong and week coupling limits, and has the the strong coupling expansion of the form
\be
h(\lambda) =  \sqrt{\frac{\lambda}{2}}  + c + \cdots
\ee
We will examine the small $v$ region of the GM solution with $Q=1$ below.
The magnon becomes gradually larger as the parameter $v$ approaches the $v\ra 0$ limit.
The dispersion relation \rf{dr} can be expanded as\footnote{Presumably the $-\sqrt{\fr{\lambda}{2}}\;  v^2$ term
should come from spacetime loops.}
\be
\Delta = \sqrt{2 \lambda}\; \Big(1 - \frac{1}{2} v^2\big) + 2 c \;\Big(1 - \frac{1}{2} v^2\Big) +
{\cal O} (v^4) \label{edr}
\ee
 The first paper in which $c$ was computed is \cite{Shenderovich:2008bs}, and $c=0$ was obtained.
Using a new summing method, the authors of \cite{Abbott:2010yb} obtained $c=-\fr{\ln 2}{2\pi}$. Both of these works employed the algebraic curve method. {  The difference in $c$ values was then attributed to different regularization methods involved in carrying out certain sums. It should be a worthwhile endeavor to see whether one of these summing methods (or possibly another) could correspond to a wellknown regularization, such as dimensional regularization, of standard QFT technique. 
 \cite{Kazakov:2004qf}\cite{Gromov:2008bz}. Below we determine the constant $c$
by using the standard worldsheet fluctuation technique and by employing dimensional regularization.}

Although the IIA GM is a relatively simple solution, the worldsheet analysis becomes much more complicated compared with, e.g., the circular or folded string configurations \cite{Frolov:2002av,Park:2005ji,Frolov:2004bh,McLoughlin:2008he,Beccaria:2010ry,Sakaguchi:2010dg,Beccaria:2010zn}. This is due to the fact that the fluctuation lagrangian around a GM solution has 2D coordinate-dependent coefficients. This feature makes determination of the eigenvalues of the kinetic operator very nontrivial.\footnote{The zero modes were found in \cite{Minahan:2007gf}. } Nevertheless, the one-loop energy shift can be computed, as we show in this paper, as a perturbative series in the parameter $v$.

\vspace{.3in}
The rest of the paper is organized as follows. In the next section, we briefly review the GM solution of \cite{Hofman:2006xt}. We start with the IIA nonlinear sigma model in a general curved IIA supergravity background. The quadratic action that results from expanding
the starting action around the GM solution has 2D coordinate-dependent coefficients.
The one-loop correction to energy is computed as a series of the parameter $v$ in section 3. We start with the bosonic sector.
The leading $v^0$ order is computed in dimensional regularization. The fermionic sector requires more care.
 The sector involves kappa symmetry fixing. We observe that a commonly used
gauge choice is at odds with the worldsheet Lorentz invariance, and the magnon solution suggests a different but natural fixing.
The coefficient $c$ turns out to be different from the algebraic results, and is given in \rf{cvalue}. In the conclusion, we
comment on the absence of $v$-linear order in accordance with \rf{edr}.
We end with summary and future directions.
\section{Quadratic action around Giant Magnon}
To set the stage for the next section where we conduct the one-loop analysis, we
briefly review the GM solution of \cite{Hofman:2006xt}. The required quadratic action
can be obtained by expanding the $AdS_4\times CP_3$ action around the GM
solution.
An important point is that the {\em supergravity} action must be in a consistent convention with the nonlinear sigma model {\em string} action. As widely known, the supergravity action can be obtained from the corresponding string action (e.g., IIA supergravity from IIA superstring) by imposing kappa symmetry on the string action in a general supergravity background. One must substitute the
$AdS_4\times CP^3$ solution of the resulting supergravity action back into the {\em original} string sigma model action. This way, the uniformity of the convention is ensured, which of course is required for the consistency of the analysis. Throughout the paper, we employ the conventions of \cite{deWit:1998tk} both for the string and supergravity action. The work of \cite{deWit:1998tk} was in the context of the membrane and 11D supergravity. For our purpose, therefore, it is necessary to reduce those theories to 10D. The reduction to IIA string was carried out in \cite{Cvetic:1999zs}.
(Several re-scalings were introduced therein in order to put the sigma model action into
 the standard form. Here we undo those re-scalings to remain within the conventions of \cite{deWit:1998tk}, i.e., to assure the use of the IIA supergravity action that comes directly out of reduction 11D supergravity.)
\subsection{review $AdS_4\times \mathbb{C}P^3$}
The supergravity action can be obtained by dimensional reduction of the membrane and 11D supergravity action given in \cite{deWit:1998tk}.
IIA supergravity admits an $AdS_4\times CP_3$ solution; it is given, in string frame, by
\begin{eqnarray}
ds^2_{IIA} &=&  ds^2_{AdS_4} +ds^2_{\mathbb{C}P^3}\\
F_{mn} &=& k\partial_{\left[ m\right.}A_{\left.n\right]}    \\
F_{(4)} &=& \frac{3}{8}kR^2\epsilon_{(4)} \\
e^{\Phi} &=& \frac{R}{k}
\end{eqnarray}
where $\epsilon_{(4)}$ is the Levi-Civita symbol on $AdS_4$, $R$ is the radius of the curvature, $\textbf{J}$ is the K$\ddot{a}$hler form on $\mathbb{C}P^3$, and $k$ is an integer-valued constant (that corresponds to the level of ABJM theory).
The metrics for the $AdS_4$ part and $\mathbb{C}P^3$ part are given respectively by
\begin{equation}
ds_{AdS_4}^2~=\fr{R^2}{4}\Big[-\cosh^2\hat{\rho} d\hat{t}^2+d\hat{\rho}^2+\sinh^2\hat{\rho}(d\hat{\theta}^2+\sin^2\hat{\theta} d\hat{\phi}^2)\Big]
\end{equation}
 and
\begin{eqnarray}
ds^2_{\mathbb{C}P^3} =&&{R^2}\Big[ d\hat{\xi}^2 +\cos^2\hat{\xi}\sin^2\hat{\xi} (d\hat{\psi}+\frac{1}{2}\cos\hat{\theta}_1 d\hat{\varphi}_1 -\frac{1}{2}\cos\hat{\theta}_2 d\hat{\varphi}_2)^2\\
&&+\frac{1}{4}\cos^2\hat{\xi} (d\hat{\theta}^2_1+\sin^2\hat{\theta}_1d\hat{\varphi}_1^2)+\frac{1}{4}\sin^2\hat{\xi} (d\hat{\theta}^2_2+\sin^2\hat{\theta}_2 d\hat{\varphi}_2^2) \Big]
\nonumber
\end{eqnarray}
with
\begin{equation}
0\leq \hat{\xi} <\frac{\pi}{2} ~,~ 0\leq \hat{\psi}<2\pi ~,~ 0\leq \hat{\theta}_i\leq\pi ~,~ 0\leq\hat{\varphi}_i<2\pi \,.
\end{equation}
In this coordinate system, the K$\ddot{a}$hler form on $\mathbb{C}P^3$ is given by $\textbf{J}=R^2dA$ where
\begin{equation}
A=\frac{1}{2} (\cos\hat{\theta}_1\cos^2\hat{\xi} d\hat{\varphi}_1 +\cos\hat{\theta}_2\sin^2\hat{\xi} d\hat{\varphi}_2 +\cos2\hat{\xi} d\hat{\psi})\, .
\end{equation}
One finds the following bosonic part of the nonlinear sigma model lagrangian:
\bea
\cL_B&=& \frac{R^2}{4} \left\{-\cosh^2\hat{\rho} (\pa\hat{t})^2+(\pa\hat{\rho})^2+\sinh^2\hat{\rho}\;[(\pa\hat{\theta})^2
           +\sin^2\hat{\theta} (\pa\hat{\phi})^2]\; \right\}  \nn\\
     && +R^2\left\{  (\pa\hat{\xi})^2 +\cos^2\hat{\xi}\sin^2\hat{\xi} \Big(\pa\hat{\psi}+\frac{1}{2}\cos\hat{\theta}_1 \pa\hat{\varphi}_1 -\frac{1}{2}\cos\hat{\theta}_2 \pa\hat{\varphi}_2\Big)^2  \right.\nn\\
&&\left.+\frac{1}{4}\cos^2\hat{\xi} \; [(\pa\hat{\theta})^2_1+\sin^2\hat{\theta}_1(\pa\hat{\varphi}_1)^2]+\frac{1}{4}\sin^2\hat{\xi} \; [(\pa\hat{\theta}_2)^2+\sin^2\hat{\theta}_2 (\pa\hat{\varphi}_2)^2]\;\right\}
\label{fullbact} \nn\\
\eea
Since $\hat{\rho} =0$ in the global coordinate system is degenerate, it is useful to use Cartesian coordinates, in which the metric takes
\begin{eqnarray}
ds^2_{AdS_4}
&=& \frac{R^2}{4} \left[-\left(\frac{1+\hat{\eta}^2}{1-\hat{\eta}^2}\right)^2 d\hat{t}^2 + \frac{4}{(1-\hat{\eta}^2)^2}\, d\vec{\hat{\eta}}\cdot d\vec{\hat{\eta}} \right]
\end{eqnarray}
These coordinates are related to the global coordinates by
\begin{equation}
\cosh\hat{\rho} ~=~ \frac{1+\hat{\eta}^2}{1-\hat{\eta}^2}
\end{equation}
and only valid for $\hat{\eta}^2=\vec{\hat{\eta}}\cdot\vec{\hat{\eta}}=\hat{\eta}^2_1+\hat{\eta}^2_2+\hat{\eta}^2_3 <1$.
{Below, the action \rf{fullbact} (with the fermionic part) is considered in the infinite conformal plane for the quantum correction computation.}
\subsection{$R\times S^2$ GM solution }
 The giant magnon solution that we consider has support in one diagonal $S^2$ sector of $AdS_4\times \mathbb{C}P^3$, and is given by\footnote{{The classical energy of the solution is (see (3.4) of \cite{Lee:2008ui})
\begin{eqnarray}
E-J 
= \sqrt{2\lambda}+v\mbox{-dependent terms}
\end{eqnarray}
which is the leading piece of \rf{edr}.}
}
\begin{eqnarray}
\hat{t} = \tau \quad \hat{\eta}_i = 0 \quad
\hat{\xi} = \frac{\pi}{4} \quad
\hat{\theta}_i = \theta_{0i} \quad
\hat{\varphi}_i = \varphi_{0i}\quad
\hat{\psi} = 0  \label{GMsol}
\end{eqnarray}
where
\begin{eqnarray}
\theta_0(x) &=& \cos^{-1}\left(\frac{1}{\gamma} \textrm{sech} x\right)\, , \nonumber\\
\varphi_0(x) &=& \t+  \tan^{-1}\left(\frac{1}{\gamma v}\tanh x\right)
\end{eqnarray}
and
\begin{equation}
x~=~\gamma(\sigma-v\t) ~~~~,~~~~ \gamma^2 ~=~ \frac{1}{1-v^2} \,,
\end{equation}
The parameter $v$ is the worldsheet velocity of the magnon.
$\th_{0i}$ and $\vf_{0i}$ have been set to $\th_0$ and $\vf_0$ respectively:
\begin{equation}
\theta_{01}=\theta_{02}\equiv \th_0 ~~~\textrm{and}~~~ \varphi_{01}=\varphi_{02}\equiv \vf_0\, .
\end{equation}
The momentum $p$ of the magnon is
related to $v$ according to
\begin{equation}
v~=~\cos\frac{p}{2} ~~~~,~~~ \gamma^{-1} ~=~\sin\frac{p}{2}\la{vp}
\end{equation}
In a generic coordinate system $\hat{X}^M$, the bosonic part of the Virasoro constraints are given by
\begin{eqnarray} \label{constraint}
G_{MN}\left(\dot{\hat{X}}^M\dot{\hat{X}}^N+\hat{X}'^{M}\hat{X}'^{N}\right) &=& 0 \\
G_{MN}\dot{\hat{X}}^M\hat{X}'^N &=&0
\end{eqnarray}
Since we will only consider the linear order of the Virasoro
constraints, the fermionic part will not contribute.
In terms of the angular coordinates introduced above, these translate to
\begin{equation}
0 = -\frac{1}{2}\dot{t} +\frac{1}{\sqrt{8}}\sin^2\theta_0 (\dot{\varphi}_0\dot{\varphi}_+ +\varphi'_0\varphi'_+) +\frac{1}{\sqrt{8}}(\dot{\theta}_0\dot{\theta}_+ +\theta'_0\theta'_+) +\frac{1}{2\sqrt{8}}\sin2\theta_0(\dot{\varphi}^2_0+\varphi'^2_0)\theta_+
\label{tdotconst}
\end{equation}
and
\begin{equation}
0 = -\frac{1}{2}t' +\frac{1}{\sqrt{8}}\sin^2\theta_0 (\dot{\varphi}_0 \varphi'_+ +\varphi'_0\dot{\varphi}_+) +\frac{1}{\sqrt{8}}(\dot{\theta}_0\theta'_+ +\theta'_0\dot{\theta}_+) +\frac{1}{\sqrt{8}}\sin 2\theta_0 \dot{\varphi}_0\varphi'_0\theta_+\, \label{tprimeconst}
\end{equation}
at the linear order in the fields. (The constraints at the zeroth order in the fields are automatically
satisfied.) Above, we have introduced
\begin{eqnarray}
\varphi_+ &\equiv & \frac{1}{\sqrt{2}} \left( \varphi_1 + \varphi_2\right)\nonumber\\
\theta_+ &\equiv & \frac{1}{\sqrt{2}} \left(\theta_1+\theta_2\right)\,
\end{eqnarray}
for convenience. The fields without " $\hat{}$ " represent the fluctuation fields. Similarly, let us define
\begin{eqnarray}
\varphi_- &\equiv & \frac{1}{\sqrt{2}} \left( \varphi_1 - \varphi_2\right)\nonumber\\
\theta_- &\equiv & \frac{1}{\sqrt{2}} \left(\theta_1-\theta_2\right)\,
\end{eqnarray}
for later use.
Note that the constraints are among $t, \vf_+, \th_+$ at this order. (All the other fluctuation fields appear at the quadratic (and higher)-order expressions for the constraints.) This, with the structure of the quadratic lagrangian, naturally divides the bosonic sector into three sub-sectors as we discuss in section 3.
\subsection{quadratic action}
Upon substituting the solution into the sigma model action and expanding the resulting action, one can show, after lengthy algebra involving re-scalings, that
\begin{eqnarray}
-4{\cal{L}}_B^{(2)} &=& (\partial t)^2 -4\sum^3_{i=1}\left[ (\partial\eta_i)^2+\eta_i^2 \right] -8(\partial\psi)^2-8(\partial\xi)^2\nonumber\\
&&-\frac{1}{2}\sin^2\th_0(\partial\varphi_+)^2
-\frac{1}{2}(\partial\theta_+)^2
\nonumber\\
&&
-\frac{1}{2}(1-2\sin^2\theta_0)(\partial\varphi_0)^2\,\theta^2_
+
-\sin2\theta_0\partial^a\varphi_0\partial_a\varphi_+\theta_+   \nonumber\\
&&-\frac{1}{2}(\partial\varphi_-)^2-\frac{1}{2}(\partial\theta_-)^2 -\frac{1}{2}\cos^2\theta_0(\partial\varphi_0)^2\,\theta^2_-\nonumber\\
&&+4\left[ \partial^a\theta_0\partial_a\theta_- +\sin^2\theta_0\partial^a\varphi_0\partial_a\varphi_- +\frac{1}{2}\sin2\theta_0(\partial\varphi_0)^2\theta_-\right]\xi \nonumber\\
&&-\frac{1}{2}\sin2\theta_0\left[ \partial^a\varphi_0\partial_a\varphi_-\theta_-\right] \nonumber\\
&&+4\sin\theta_0\partial^a\varphi_0\partial_a\psi\, \theta_- -4\cos\theta_0\partial^a\varphi_-\partial_a\psi \, .
  \label{LBq}
\end{eqnarray}
for the bosonic part, and
\begin{equation}
-4{\cal{L}}_F ~=~ 4e^{\Phi/3}\bar{\Theta}(\eta^{ab}+\epsilon^{ab}\Gamma_{11})e_a \left[\left(\partial_b+\frac{1}{4}w_b\right) +\Gamma\cdot F e_b\right]\Theta
\label{LFq}
\end{equation}
for the fermionic part. Above,
 \bea
 \bar{\Theta}\equiv \Theta^\dag\Gamma^0,\quad \epsilon^{\tau\sigma}= 1, \quad e_a\equiv \partial_aX^Me_M^{\phantom{M}A}\Gamma_A,\quad w_a\equiv\partial_aX^Mw_M^{\phantom{M}AB}\Gamma_{AB}
 \eea
  and
\begin{equation}
\Gamma\cdot F \equiv \frac{1}{8}e^\phi(-\Gamma_{11}\Gamma\cdot F_2+\Gamma\cdot F_4)
\equiv \frac{1}{8}e^\phi\Big[-\fr12\Gamma_{11}\Gamma^{AB} (F_2)_{AB}+\fr1{4!}\Gamma^{ABCD} (F_4)_{ABCD}\Big]
\la{gf}
\end{equation}
Since the fermionic action is already quadratic in $\Th$, it is only necessary to keep
 the leading-order (i.e., zeroth order) terms when evaluating the expressions above such as \rf{gf}.
Substituting the solution, one can show, in particular, that
\bea
\G\cdot F_2 &=& \fr{2k}{R^2}(\G^{45}+\G^{67}+\G^{89}) \nn\\
  \G\cdot F_4 &=& \fr{6k}{R^2}\G^{0123}\;\; \text{with}\; \ve_{0123}=1,
\eea
These results will be used in the next section where the worldsheet one-loop energy
correction is analyzed. {Although the Lagrangian \rf{LBq} and \rf{LFq} have coordinate-dependent coefficients, they are worldsheet coordinates but not spacetime coordinates and energy is well-defined. (Also, we use the conformal gauge not an axial type gauge where the worldsheet time is identified with the time coordinate of the spacetime.) There might be a field redefinition that can simplify the analysis of the following section. We adopt the method that we have adopted, in spite of being of brute-force, because it can be applied in cases where
such a field definition cannot easily be obtained.}
\section{One-loop energy correction in $v$-series}
Let us compute the one-loop energy shift for the magnon.\footnote{The AdS$_5 \times$S$_5$ case was considered  in \cite{Papathanasiou:2007gd}. Some of the related works in AdS$_4 \times$CP$_3$ include \cite{Lukowski:2008eq,Abbott:2009um,Bandres:2009kw,Ahn:2010eg,Astolfi:2011ju,Astolfi:2011bg,Forini:2012bb}.}
Although we are dealing with the path integral that is quadratic in the fields,
the coordinate dependence of the coefficients and nontrivial couplings between the fields
make the full evaluation of the one-loop energy nontrivial. The analysis becomes more manageable once the lagrangian
is expanded in terms of the small $v$, the velocity parameter that appears in the magnon
solution.
Let us start with generalities, and consider
\bea
\int e^{i\int \cL_0+v\cL_1+v^2\cL_2+\cdots}
\eea
where $\cL_0$ is the $v$-independent part; it is non-diagonal, and has coefficients that are functions of 2D space coordinates.
 $\cL_1$ ($\cL_2$) are the
terms in linear (quadratic) order in $v$.
Since the kinetic operator associated with $\cL_0$ has position-dependent coefficients and the fields
are coupled in nontrivial ways, explicit determination of the propagator does not seem straightforward. For this reason, we first put the system in a 2D box of length $L$ and go to the discrete 2D momentum space.\footnote{Putting a system in a box - which regulates the infrared divergences - and adopting a ultraviolet regulator is a common practice in the kink literature \cite{Goldhaber:2004kn}.}
After evaluating the one-loop energy, we convert the resulting expressions into continuous momentum space by taking the appropriate continuum limit.

\ssk
Let us put the system in a 2D box of length $L$ for each side, and consider the expansion of fields in terms
of the following complete set, $\{e^{i\vec{p}_n\cdot \vec{z}}\}$, with
\bea
  \vec{p}_n=(p_{n\t},p_{n\s})=\fr{2\pi }{L}\vn=\fr{2\pi }{L}( n_\t, n_\s),\quad n_\s,n_\t=0,\pm 1,....,\pm \infty
\eea
 The set satisfies the usual orthonormality condition
\bea
  \int_{-L/2}^{L/2} \int_{-L/2}^{L/2} \fr{d^2z}{L^2}\; e^{-i(\vec{p}_m-\vec{p}_n)\cdot \vec{z}}=\d_{m-n,0}
  \label{oth}
\eea
The actual analysis in section \ref{bp} reveals that the Fourier-transformed kinetic parts
are diagonal after the $L\ra \infty$ limit is taken.
Let us consider the bosonic sector, and collectively denote
the bosonic fields by $\Phi$:
\bea
\Phi:\; \mbox{a collective representation for the bosonic fields}
\eea
The analysis for the fermionic sector is the same except for the usual sign change.
Let us introduce the Fourier expansion
\bea
\Phi=\dfrac{1}{L}\sum_{\vec{n}} e^{i\vec{p}_{n}\cdot \vec{z}}
\Phit_{\vn} \quad \mbox{with}\quad \Phit_{\vec{n}}^\dagger=\Phit_{-\vec{n}}
\eea
where $\vec{z}$ denotes the 2D worldsheet coordinates $(\t,\s)$,
\bea
\vec{z}=(\t,\s)
\eea
 and the second equation originates from the reality of $\Phi$. After adding the source terms, the $v^0$-order action in the momentum space takes the following schematic form (the summation convention is understood)
\bea
- \Phit_{\vm} M_{\vm\vn} \Phit_{\vn} + \Jt_{\vn}^\dagger \Phit_{\vn}+ \Phit_{\vn}^\dagger \Jt_{\vn}
\eea
where $M_{mn}, J_n$ are the Fourier transformations of kinetic operator and the source term. The one-loop energy at $v^0$-order comes from
$M_{mn}$. Although the focus of this work is the $v^0$-order, we present the general expression for the $v$-order
energy for future purpose.
The vertex terms (i.e., $v$-dependent terms) $\int \Phi V_{\Phi\Phi}\Phi$ go as
\bea
\int \fr{d^2z}{L^2}\;\Phi V_{\Phi\Phi}\Phi &=& \fr1{L^2}\sum_{\vn,\vm}\int \fr{d^2z}{L^2}\;\Phit_{\vm}^\dagger
e^{-i\vec{p}_{m}\cdot \vec{z}}  V_{\Phi\Phi} \Phit_{\vn} e^{i\vec{p}_{\vn}\cdot \vec{z}}
\equiv \Phit_{\vm}^\dagger (\Vt_{\Phi\Phi})_{\vm\vn} \Phit_{\vn}  \nn\\
\eea
The matrix $(\Vt_{\Phi\Phi})_{\vm\vn}$ is the Fourier transformation of $\Vt_{\Phi\Phi}$,
\bea
(\Vt_{\Phi\Phi})_{\vm\vn} &\equiv & \int \fr{d^2z}{L^2}\;e^{-i\vec{p}_{m}\cdot \vec{z}}  V_{\Phi\Phi}\;
e^{i\vec{p}_n\cdot \vec{z}}
\eea
At the $v$-order, this leads to the following schematic expression for the one-loop energy:
\bea
  (\Vt_{\Phi\Phi})_{\vm\vn} M^{-1}_{\vn\vm} \label{VMinv}
\eea
where we keep the linear order terms in $V_{\Phi\Phi}$. The inverse in $M^{-1}$ should be taken in the tensor product space of the 4 by 4 matrix space and
the space of the $(m,n)$ indices. This would be a highly memory-demanding procedure in Mathematica computation. Fortunately, however, the matrix $M$
turns out to be diagonal in the $(m,n)$ space in the large $L$ limit, and the inverse-taking procedure becomes simple.
\subsection{bosonic part   \la{bp}}
We compute the one-loop energy by adopting conformal gauge (which we discuss in sec \ref{3b3}).
The bosonic part can be divided into three sectors, the $\eta$ sector, $(\psi,\xi,\vf_-,\th_-)$ sector and $(t,\th_+, \vf_+)$ sector. The Virasoro constraints pertain to $(t,\th_+, \vf_+)$ sector.
\subsubsection{$\eta$ sector}
The $\eta$ sector of the lagrangian takes a simple form with constant
coefficients, and one can simply adopt the usual approach
of computing the one-loop energy. However, we take this sector to illustrate the procedure that we will use heavily
in the other sectors, and demonstrate that the procedure reproduces the standard result when applied to
a case with constant coefficients. While doing so, we find proper normalizations as well.
Putting the system in a box with area $L^2$
{
\begin{eqnarray}
\int(-4){\cal{L}}_{\eta}^{(2)} &\Rightarrow &
(-4R^2)\int \left[ (\partial\eta_i)^2+\eta_i^2 \right]\nn\\
&=& (-4R^2)\fr1{L^2}\sum^3_{i=1}\sum_{\vec{n}}\tilde{\eta}_i^{\vn \dagger}(\vec{p}_n^{\,2}+1)\tilde{\eta}_i^{\vn}
\eea
}
where we have used the proper normalization in which the sums over
$m,n$ come with $\fr1{\sqrt{L^2}}$.
The path integral over $d\eta_i^n$ measure yields the following contribution
\bea
-\sum_{\vec{n}}\ln (\vec{p}_n^{\,2}+1)
\eea
for the one-loop energy. (The overall coefficient is irrelevant, not being recorded accurately.) After taking the continuum limit according to
\bea
\vec{p}_n \ra \vec{p}\quad ,\quad  \fr{1}{L^2}\sum_{\vec{n}} \ra \int d^2p
\eea
one obtains the standard expression
\bea
-3\int d^2p\; \ln (\vec{p}^{\,2}+1)
\eea
where the factor 3 came from $\sum_i^3$.
\subsubsection{$(\psi,\xi,\vf_-,\th_-)$ sector}
The lagrangian for this sector is given by
\begin{eqnarray}
[-4{\cal{L}}_{(\psi,\xi,\vf_-,\th_-)}^{(2)}] &=& -8(\partial\psi)^2-8(\partial\xi)^2\nonumber\\
&&-\frac{1}{2}(\partial\varphi_-)^2-\frac{1}{2}(\partial\theta_-)^2 -\frac{1}{2}\cos^2\theta_0(\partial\varphi_0)^2\,\theta^2_-\nonumber\\
&&+4\left[ \partial^a\theta_0\partial_a\theta_- +\sin^2\theta_0\partial^a\varphi_0\partial_a\varphi_- +\frac{1}{2}\sin2\theta_0(\partial\varphi_0)^2\theta_-\right]\xi \nonumber\\
&&-\frac{1}{2}\sin2\theta_0\left[ \partial^a\varphi_0\partial_a\varphi_-\theta_-\right] \nonumber\\
&&+4\sin\theta_0\partial^a\varphi_0\partial_a\psi\, \theta_- -4\cos\theta_0\partial^a\varphi_-\partial_a\psi  \la{44L}
\end{eqnarray}
  In terms of the Fourier modes, the action
  can be rewritten as
\bea
\fr{1}{L^2}\sum_{\vn,\vm}\int \fr{d^2z}{L^2} e^{i(p_{n}-p_{m})\cdot z}(\psit^{m \dagger} \; \xit^{m \dagger} \;\vft_-^{m \dagger} \; \tht_-^{m \dagger} )M_{4\times 4} \left(\psit^{n}\,\xit^{n}\,\vft_-^{n}\,\tht_-^{n}\right)^T
\eea
where $M_{4\times 4}$ is a  4 by 4 matrix given by
\[
               -8p_{n}^2   \quad,\quad    0 \;,\;   -2\cos\th_0 p_{m}^a p_{a n}  \quad,\quad -2i\sin\th_0 \pa_a \vf_0 p_{n}^a-\pa_a(\sin\th_0\, \pa^a \vf_0)
       \]
 \[
                  0     \quad,\quad    -8p_{n}^2  \quad,\quad 2i\sin^2\th_0 \pa_a\vf_0 p_{m}^a-\pa^a(\sin^2\th_0 \pa_a\vf_0)   \quad,\quad 2i\pa_a\th_0 p_{m}^a-\pa^a\pa_a \th_0+  \sin2\th_0 (\pa\vf_0)^2 \]
\[          \!\!\!\!\!\!\!\!\!\!\!\!\!\!\!\!
             -2\cos\th_0 p_{m}^a p_{a n} \;,\;  -2i\sin^2\th_0 \pa_a\vf_0 p_{m}^a -\pa^a(\sin^2\th_0 \pa_a\vf_0) \;,\;  -\fr12p_{n}^2  \;,\; \fr{ip_m^a}4\sin 2\th_0 (\pa_a\vf_0)+\fr18 \pa_a(\sin2\th_0 \pa^a \vf_0)
                \]
\bea
     && \!\!\!\!\!\!\!\!\!\!\!\!\!\!\!\!\!\!\!\!\!\!\!\!\!\!\!\!
     \!\!\!\!\!\!\!\!\!\!\!\!\!\!\!\!\!\!\!\!
      2i\sin\th_0 \pa_a \vf_0 p_{n}^a-\pa_a(\sin\th_0\, \pa^a \vf_0) \;,\;  -2i\pa_a\th_0 p_{m}^a-\pa^a\pa_a \th_0+  \sin2\th_0 (\pa\vf_0)^2 \;,\;   -\fr{ip_m^a}4\sin 2\th_0 (\pa_a\vf_0)+\fr18 \pa_a(\sin2\th_0 \pa^a \vf_0)  \nn\\
       && \hspace{3in},\quad   -\fr12\Big(p_{n}^2+ \cos^2\th_0 (\pa\vf_0)^2\Big)
\eea
 We made heavy use of Mathematica in the following analysis.
It is convenient to split the sum over $(\vec{n},\vec{m})$ into four sectors. Defining
$\vec{n}=(n_\t,n_\s)$, $\vec{m}=(m_\t,m_\s)$, they are
\bea
(n_\t\neq m_\t,n_\s\neq m_\s),\quad (n_\t= m_\t,n_\s= m_\s),\quad (n_\t\neq m_\t,n_\s=m_\s),\quad (n_\t= m_\t,n_\s\neq m_\s)
\nn\\
\eea
\underline {At the $v^0$-order}, the nonzero contributions come from $(n_\t=m_\t,n_\s= m_\s)$ and $(n_\t=m_\t,n_\s\neq m_\s)$ sectors.
After taking $L\ra \infty$ limit in
 the $(n_\t=m_\t,n_\s\neq m_\s)$ sector, Mathematica produces
an expression that is antisymmetric in the tensor product space of the 4 by 4 matrix and the $(\vn,\vm)$ space, which
therefore, does not contribute to the one-loop correction.
For the $(n_\t=m_\t,n_\s=m_\s)$ sector, one finds
\bea
R^2 diag(-8\vec{p}_n^{\,2},-8\vec{p}_n^{\,2},-\fr12 \vec{p}_n^{\,2},-\fr12 \vec{p}_n^{\,2}) \la{4by4vzero}
\eea
As a matter of fact, some of the off-diagonal entries survive after Fourier transformation; they are
irrelevant for the one-loop energy. This is because the off-diagonal part is antisymmetric and therefore
is removed by taking trace in the one-loop computation. Let us illustrate how all of the off-diagonal entries vanish with the (1,3) entry that contains $\cos\th_0$
in the $(n_\t=m_\t,n_\s = m_\s)$ sector. The integral produces
\bea
 \frac{4}{L} \text{arctan}\left(\text{tanh}\left(\frac{L}{4}\right)\right)
\eea
One can show that this expression vanishes in the large-$L$ limit.
Performing the path integral over $\left(\psit^{n}\,\xit^{n}\,\vft_-^{n}\,\tht_-^{n}\right)$ and
their conjugates and going to the continuum limit, the relevant part of the one-loop energy contribution from (\ref{4by4vzero})
is given by
\bea
-4 \int d^2p \ln p^2
\eea
\subsubsection{$(t,\th_+, \vf_+)$ sector \la{3b3}}
\ni The $(t,\th_+, \vf_+)$ sector of the
bosonic action is
\begin{eqnarray}
[-4{\cal{L}}_B ]_{t,\th_+, \vf_+} &=&  (\partial t)^2
-\frac{1}{2}(\partial\theta_+)^2
-\frac{1}{2}(1-2\sin^2\theta_0)(\partial\varphi_0)^2\,\theta^2_+ \nn\\
&& -\sin2\theta_0\partial^a\varphi_0\partial_a\varphi_+\theta_+
-\frac{1}{2}\sin^2\th_0(\partial\varphi_+)^2
\label{3by3}
\eea
For this sector, the Virasoro constraints \rf{tdotconst} and \rf{tprimeconst} must be
taken into account.
Using the residual symmetry, one can impose
\bea
t=0
\eea
The action \rf{3by3} then becomes
\begin{eqnarray}
[-4{\cal{L}}_B ]_{\th_+, \vf_+} &=&
-\frac{1}{2}(\partial\theta_+)^2
-\frac{1}{2}(1-2\sin^2\theta_0)(\partial\varphi_0)^2\,\theta^2_+ \nn\\
&& -\sin2\theta_0\partial^a\varphi_0\partial_a\varphi_+\theta_+
-\frac{1}{2}\sin^2\th_0(\partial\varphi_+)^2
\label{3by3r}
\eea
The Virasoro constraints now take
\bea
&&\sin^2\theta_0 (\dot{\varphi}_0\dot{\varphi}_+ +\varphi'_0\varphi'_+) +(\dot{\theta}_0\dot{\theta}_+ +\theta'_0\theta'_+) +\frac{1}{2}\sin2\theta_0(\dot{\varphi}^2_0+\varphi'^2_0)\theta_+=0\nn\\
&&\sin^2\theta_0 (\dot{\varphi}_0 \varphi'_+ +\varphi'_0\dot{\varphi}_+) +(\dot{\theta}_0\theta'_+ +\theta'_0\dot{\theta}_+) +\sin 2\theta_0 \dot{\varphi}_0\varphi'_0\theta_+=0 \label{tprimeconstr}
\eea
These can easily be solved for $\dot{\vf}_+,\vf_+' $, and the solutions can be substituted into \rf{3by3r}.
The rest of the analysis is similar to that of the previous section; the one-loop contribution from this sector is
\bea
-\int d^2p\;\ln (p^2+1)
\eea
\subsection{fermionic part  \la{fp}}
Let us quote the quadratic fermionic action for convenience:
\begin{equation}
-4{\cal{L}}_F ~=~ 4e^{\Phi/3}\bar{\Theta}(\eta^{ab}+\epsilon^{ab}\Gamma_{11})e_a \left[\left(\partial_b+\frac{1}{4}w_b\right) +\Gamma\cdot F e_b\right]\Theta
\label{resfa}
\end{equation}
The fermionic coordinates are such that $\Th= \Th^1+\Th^2$ with
\bea
\Theta^1 &\equiv& \left(\begin{array}{c}
               \varTheta^1  \\
               0
             \end{array}\right)\quad,\quad
\Theta^2 \equiv \left(\begin{array}{c}
               0       \\
               \varTheta^2
             \end{array}\right)
\eea
The matrix that appears in \rf{resfa} turns out to be half-ranked when the magnon solution
is substituted. In other words, half of the components in $\Th$ do not appear when the action \rf{resfa}
is expanded. This naturally suggest a gauge fixing in which one sets those components to zero.\footnote{To our surprise, a certain kappa-symmetry gauge choice
 seems incompatible (at least apparently) with the worldsheet Lorentz invariance.
  We illustrate this point with an innocuous-looking
 gauge choice $(1-\G^{11})\Th=0$.
 It is easier to see the issue in flat space. A convenient choice for $\G^{11}$ is
\bea
\G^{11} &=& \left(\begin{array}{cc}
               1  & 0\\
               0      & -1
             \end{array}\right)
\eea
The $\Th^1, \Th^2$ field equations take
\bea
(-\pa_\t+\pa_\s)\Th^1=0\quad,\quad (\pa_\t+\pa_\s)\Th^2=0
\eea
The particular gauge choice removes the entire $\Th^2$ (i.e., the left-moving modes) thereby breaking the
2D Lorentz invariance of the fermionic part of the Virasoro constraints, i.e., the fermionic part that is not explicitly recorded in \rf{tdotconst} and \rf{tprimeconst}. Although it might be possible to restore the 2D invariance at the end, it might take some complicated steps.
If $\k$-fixing were necessary, the following gauge fixing would be a good choice for example:
\bea
(\G^0+\G^1)\Th=0 \la{newgauge}
\eea
}
Consider the following Fourier transformation
\bea
\Th =\sum_n e^{i\vec{p}_n\cdot \vec{z}} \tilde{\Th}_n
\eea
In the Fourier transformed space, one can separated the momentum-dependent part of the
matrix in \rf{resfa} from the momentum-independent part. It turns out that the momentum independent part
vanishes by symmetric sum.
One subtle issue concerns how to take hermitian conjugation of the kinetic matrix in the $\vn$-vector space. It turns out that the
correct hermitian conjugation in the momentum space involves $\vn\ra -\vn$ as well.\footnote{This subtlety should be attributed to the fact that both positive and negative integers were used to label the matrix in the $\vn$-vector space. }
To see this in more detail, let us consider one of the momentum independent entries,
\bea
-\fr{R}{16}\sum_{-\infty}^\infty \Th_{\vn}^{1\dagger}  \Th_{\vn}^4
\eea
where $1$ and $4$ are the spinor indices, and impose the hermiticity requirement by using the
the usual hermitian conjugation (, i.e., the one that does not involve $\vn\ra -\vn$,)
\bea
&& \sum_{n=-\infty}^\infty \Th_{\vn}^{1\dagger}  \Th_{\vn}^4=
\Big(\sum_{n=-\infty}^\infty \Th_{\vn}^{1\dagger}  \Th_{\vn}^4\Big)^\dagger
= \sum_{n=-\infty}^\infty \Th_{\vn}^{4\dagger}  \Th_{\vn}^1   \nn\\
=&& -\sum_{n=-\infty}^\infty \Th_{\vn}^{1}  \Th_{\vn}^{4*}
=-\sum_{n=-\infty}^\infty \Th_{\vn}^{1}  \Th_{-{\vn}}^{4}
=-\sum_{n=-\infty}^\infty \Th_{-{\vn}}^{1\dagger}  \Th_{-{\vn}}^{4}\nn\\
=&& -\sum_{n=-\infty}^\infty \Th_{{\vn}}^{1\dagger}  \Th_{{\vn}}^{4}
\eea
Therefore, the usual hermitian conjugation,
lead to
\bea
&& \sum_{n=-\infty}^\infty \Th_{\vn}^{1\dagger}  \Th_{\vn}^4=0
\eea
As a matter of fact even if one uses the hermitian conjugation that {\em does} involve
 $\vn\ra -\vn$, one gets the same result. However, those two conjugations lead to different results for momentum dependent terms to which we now turn; for example consider
 \bea
 \fr{iR}{2\sqrt{2}}\sum_{n=-\infty}^\infty\Th_{\vn}^{1\dagger}(p_{n_\t}+p_{n_\s})\Th_{\vn}^9
 \eea
The usual hermitian conjugation yields
 \bea
&& i\sum_{n=-\infty}^\infty\Th_{\vn}^{1\dagger}(p_{n_\t}+p_{n_\s})\Th_{\vn}^9
 =\Big(i\sum_{n=-\infty}^\infty\Th_{\vn}^{1\dagger}(p_{n_\t}+p_{n_\s})\Th_{\vn}^9\Big)^\dagger \nn\\
=&& -i\sum_{n=-\infty}^\infty\Th_{\vn}^{9\dagger}(p_{n_\t}+p_{n_\s})\Th_{\vn}^1
=i\sum_{n=-\infty}^\infty\Th_{\vn}^{1}(p_{n_\t}+p_{n_\s})\Th_{\vn}^{9*}\nn\\
=&& i\sum_{n=-\infty}^\infty\Th_{\vn}^{1}(p_{n_\t}+p_{n_\s})\Th_{-\vn}^{9}
= i\sum_{n=-\infty}^\infty\Th_{-\vn}^{1\dagger}(p_{n_\t}+p_{n_\s})\Th_{-\vn}^{9}\nn\\
=&& -i\sum_{n=-\infty}^\infty\Th_{\vn}^{1\dagger}(p_{n_\t}+p_{n_\s})\Th_{\vn}^{9}  \la{fher}
 \eea
Therefore one gets
\bea
 i\sum_{n=-\infty}^\infty\Th_{\vn}^{1\dagger}(p_{n_\t}+p_{n_\s})\Th_{\vn}^9=0
\eea
Because of this, the entire fermionic matrix vanishes, and this cannot be true.
If one uses  the hermitian conjugation that does involve
 $\vn\ra -\vn$, additional minus sign appears in the far right hand side of \rf{fher}, making the result non-vanishing.\footnote{Another indication towards the correctness of the hermitian conjugation
 is that without it some of the spurious modes survive rendering the integrand of the 2D momentum space non-covariant.}
\ssk
\ni \underline{At the $v^0$-order}, the relevant part of one-loop contribution turns out to be
\bea
8\int d^2p\;\ln p^2
\eea
\subsection{Combining bosonic and fermionic contributions}
Let us combine the bosonic and fermionic results.
The total energy $v^0$ order is given by
\bea
&& \!\!\!-\int d^2 p\;\Big(4\ln(p^2+1)+4\ln p^2-8\ln p^2\Big)  \nn\\
=&&\!\!\!-\int d^2 p\; \big(4\ln(p^2+1)-4\ln p^2\Big)
  \la{esum}
\eea
The first term in the first line is a combination of the results from the $\eta$-sector and $\th_+$-sector while
the second term comes from the 4 by 4 sector. The third term comes from the fermionic sector.
The $\int d^2 p\;4\ln(p^2+1)$ can be evaluated as follows. Let us define
\bea
&& \d E_{(\eta,\th_+)}=-4\int d^2p\;\ln {(\vec{p}^{\, 2}+m^2)}
\eea
where we have introduced a ``mass" parameter, $m=1$.
In the dimensional regularization that we have adopted, this can be evaluated as follows\footnote{Alternatively, 
\bea
&& \d E_{(\eta,\th_+)}=-4\int d^2p\;\ln {(\vec{p}^{\, 2}+m^2)}
=\left[4 \fr{\pa}{\pa u}\int d^2p\;\fr1{(\vec{p}^{\, 2}+m^2)^u}\right]_{u=0}\nn\\
&&=\left[4 \fr{\pa}{\pa u}\fr1{4\pi}\fr{\G(u-1)}{\G(u)(m^{2u-2})}\right]_{u=0}
=\fr1{\pi}\fr{\G(-1)}{m^{-2}}=-\fr1{\pi}\fr{\G(0)}{m^{-2}}
\eea
which yields the same result as \rf{Eshift}.
}:
\bea
\fr{\pa}{\pa m^2}\d E_{(\eta,\th_+)}=-4\int d^2p\;\fr1{\vec{p}^{\,2}+m^2}=-4\fr{\G(0)}{4\pi}
\eea
{where $\G$ is gamma function.}
This implies
\bea
\d E_{(\eta,\th_+)}=-4\fr{\G(0)}{4\pi}m^2+C=-\fr{\G(0)}{\pi}+C
\eea
where we have used $m^2=1$ for the second equality. The constant $C$ gets cancelled by
the $\ln p^2$ term in \rf{esum} and the sought-for one-loop energy shift $\d E$ is 
given, {  up to the issue of renormalization conditions}, by
\bea
\d E =-\fr{1}{\pi}  \la{Eshift}
\eea
{  after absorbing $\G(0)$ by wave function renormalization. }
 
\section{Conclusion}
In this work, we have analyzed the worldsheet one-loop energy shift in dimensional regularization. The presence of the volume factor $\G(0)$ is a general feature of loop corrections, and should absorbed by a renormalization procedure; we have absorbed it by wave function renormalization. (Renormalization was considered in \cite{Roiban:2007jf} in the context of a folded rotating string.) Presumably there should be the corresponding regularization and renormalization procedure in the dual theory.
Our analysis has led to the following value of the coefficient $c$ that appears in \rf{edr}\footnote{{This will require, e.g., wave-function renormalization in which the volume factor is absorbed.
 It is necessary to go to higher orders to check the (in)consistency of the renormalization scheme against eq.\rf{dr}. Systematic of renormalization will not be pursued further in this work.}},
\bea
c=-\fr1{\pi}  \la{cvalue}
\eea
It will be worthwhile to check whether the $v^0$-order propagator could be determined analytically, and cross-check
various results obtained in this work. Presumably, clever field redefinitions would be required to that end.

{ The result \rf{cvalue} is different (at least on face value) from the $c$ value obtained in 
\cite{Shenderovich:2008bs} or \cite{Abbott:2010yb}. We carefully checked computations multiple times and do not believe that there is any error; perhaps the discrepancy should be attributed to the regularization methods. It is well known that not all regularizations lead to the same result, even for a physical quantity, in general, and a regularization should be viewed as part of the definition of the theory. The regularization methods used in \cite{Shenderovich:2008bs} or in \cite{Abbott:2010yb} should not correspond to dimensional regularization of QFT up to issues such as counter terms and renormalization conditions. It would also be worthwhile to try to come up with a renormalization scheme and renormalization conditions of the dual theory that reproduce the result at the given order in this work. The real test then will be checking the next order.}

One obvious future direction is to compute the $v^1$ and $v^2$-order energy shifts, and check the result
against \rf{edr}.
As a matter of fact, we have carried out some preliminary calculations at the $v$ order, and the result seems to indicate
a vanishing outcome in accordance with \rf{edr}. We plan to report
on further clarification on $v$ and $v^2$ orders in the near future.
Finally, it would be useful to repeat computations by employing the formulations
of \cite{Arutyunov:2008if}\cite{Stefanski:2008ik}\cite{Gomis:2008jt}. These formulations
have an advantage of the manifest AdS$_4\times$CP$_3$ isometry.
\vskip 1cm
\ni{\bf Acknowledgments}\\
\ni We thank Lata Kharkwal, Siyoung Nam, Chanyong Park and Pichai Ramadevi for their collaborations at various stages of the project. We also thank for Arkardy Tseytlin for pointing out various references. 
\newpage
\appendix
\newpage
\newpage

\end{document}